\documentclass[letterpaper, 10 pt, conference]{ieeeconf} 
\IEEEoverridecommandlockouts

\usepackage{amsthm}
\usepackage{mathtools}
\usepackage{amssymb}
\usepackage[normalem]{ulem}

\usepackage{amsmath}

\usepackage{bm}
\usepackage{nomencl}
\usepackage{subfig}
\usepackage{cancel}

\usepackage[switch,mathlines]{lineno}
\usepackage{soul}
\usepackage{xfrac}
\usepackage{bbm}
\usepackage{breqn}
\usepackage{graphicx}
\usepackage{cite}
\usepackage{multirow}
\usepackage{inputenc}
\usepackage{enumerate} 
\usepackage{textcomp}
\usepackage{listings}
\usepackage{array}
\usepackage{tikz}
\usepackage{url}
\usepackage{algpseudocode}
\usepackage{algorithm}
\makenomenclature

\algnewcommand{\Inputs}[1]{%
 \State \textbf{Inputs:}
 \Statex \hspace*{\algorithmicindent}\parbox[t]{.8\linewidth}{\raggedright #1}
}
\algnewcommand{\Initialize}[1]{%
 \State \textbf{Initialize:}
 \Statex \hspace*{\algorithmicindent}\parbox[t]{.8\linewidth}{\raggedright #1}
}

\usepackage{stfloats}
\usepackage[hidelinks]{hyperref}

\newtheorem{theorem}{Theorem}

\theoremstyle{definition} 

\theoremstyle{remark} 
\newtheorem{remark}{Remark}
\newtheorem{example}{Example}

\title{\LARGE \bf
Statistical analysis and method to quantify the impact of measurement uncertainty on dynamic mode decomposition} 
\author{Pooja Algikar, Pranav Sharma, Marcos Netto, and Lamine Mili
\thanks{P. Algikar and L. Mili are with the Bradley Department of Electrical and Computer Engineering, Virginia Polytechnic Institute and State University, VA 22043, USA. P. Sharma is with the National Renewable Energy Laboratory, Golden, CO 80401, USA. M. Netto is with the Helen and John C. Hartmann Department of Electrical and Computer Engineering, New Jersey Institute of Technology, Newark, NJ 07102, USA.}
\thanks{This work was authored in part by the National Renewable Energy Laboratory, operated by Alliance for Sustainable Energy, LLC, for the U.S. Department of Energy (DOE) under Contract No. DE-AC36-08GO28308. Funding provided by U.S. Department of Energy Office of Electricity. The views expressed in the article do not necessarily represent the views of the DOE or the U.S. Government. The U.S. Government and the publisher, by accepting the article for publication, acknowledge that the U.S. Government retains a nonexclusive, paid-up, irrevocable, worldwide license to publish or reproduce the published form of this work or allow others to do so, for U.S. Government purposes.}
}

\begin{document}

\maketitle
\thispagestyle{empty}
\pagestyle{empty}


\begin{abstract}
We apply random matrix theory to study the impact of measurement uncertainty on dynamic mode decomposition. Specifically, when the measurements follow a normal probability density function, we show how the moments of that density propagate through the dynamic mode decomposition. While we focus on the first and second moments, the analytical expressions we derive are general and can be extended to higher-order moments. Furthermore, the proposed numerical method for propagating uncertainty is agnostic of specific dynamic mode decomposition formulations. Of particular relevance, the estimated second moments provide confidence bounds that may be used as a metric of \emph{trustworthiness}, that is, how much one can rely on a finite-dimensional linear operator to represent an underlying dynamical system. We perform numerical experiments on two canonical systems and verify the estimated confidence levels by comparing the moments with those obtained from Monte Carlo simulations.
\end{abstract}

\section{Introduction}

Real-world systems are complex and nonlinear. Identifying unknown dynamics from data and learning intrinsic coordinates that enable a linear representation of the underlying nonlinear dynamics are two of the most pressing goals of modern dynamical systems \cite{brunton2022data}. The Koopman operator approach has found tremendous success in these goals \cite{Mauroy2020}. In simple words, the Koopman operator transforms a finite-dimensional nonlinear dynamical system into an infinite-dimensional linear system. This transformation enables the use of well-developed principles and tools in linear algebra to study nonlinear dynamical systems without neglecting nonlinearities \cite{brunton2019notes}. Dynamic mode decomposition (DMD) \cite{Schmid2010}, one of the numerical methods associated with the Koopman operator \cite{Rowley2009}, has been widely adopted.
{Indeed, DMD has been applied to robotic systems \cite{Bruder2021}, traffic control systems \cite{Avila2020}, power grids \cite{sharma2021data}, and other engineering systems where sensor measurements are used for real-time operations and control. For example, in power grids, data obtained from instruments called phasor measurement units can be used to assess the stability of the system without having to solve nonlinear differential equations \cite{Susuki2014}, identify groups of synchronous generators oscillating coherently after a disturbance \cite{Susuki2011}, and quantify the interplay between state variables and oscillatory modes \cite{Netto2019}}. 

Like any data-driven method, however, DMD is susceptible to the quality of the measurements \cite{duke2012error}. The noise and uncertainty associated with the measurements directly impact the accuracy of analytical conclusions and control decisions.
Therefore, when the DMD method is applied to experimental or field data, one must account for the inherent uncertainty associated with measurements. N{\"u}ske et al. \cite{nuske2023finite} derived probabilistic bounds on the extended dynamic mode decomposition \cite{Williams2015ADecomposition}. They examine systems described by ordinary and stochastic differential equations from an ergodic and independent identically distributed sample perspective. To formulate theoretical constraints on error, the authors split the variance of the random matrix that forms the basis functions into the terms associated with the asymptotic contribution and the number of data points. Interestingly, the approximation error is quantified using the Frobenius norm between the Galerkin projection of the Koopman generator onto the space of observables and the operator computed from finite samples. The Frobenius norm measures the overall difference between the two representations. However, its insensitivity to structural errors may limit the ability to understand local uncertainty.

Colbrook et al. \cite{colbrook2024rigorous} enhanced the reliability of dynamic mode decomposition by showing a rigorous convergence on the spectral information of Koopman operators from trajectory data for chaotic systems. Zhang and Zuazua \cite{zhang2023quantitative} set boundaries on the approximation error of the extended DMD method but do not account for measurement noise. They determine convergence in terms of the largest and smallest eigenvalues of the stiffness matrix. They approximate the stiffness matrix that constitutes the dictionary functions with a normal probability distribution. Dawson et al. \cite{Dawson2014CharacterizingDecomposition} introduced a formulation for sensor noise as a random matrix. The bias induced by additive noise levels is numerically quantified through the expectation of the DMD operator. This approximation holds only for small noise levels. Finally, the bias correction is proposed by downweighting the expectation of the DMD operator based on the variance of the noise levels.

{Exact quantitative uncertainty analysis is needed to mitigate the risks associated with identifying linear approximations of nonlinear dynamics using measurements.}
We propose an uncertainty quantification method based on the random matrix formulation of DMD, where each of its constituent elements is treated as a random variable tasked with deducing its first and second moments. The analytical expressions of the exact estimation of these first and second moments are developed. These second moments model the uncertainty propagated from the system measurements through the algorithmic steps of the DMD method. With detailed quantitative uncertainty analysis, the proposed method enables informed decision-making under uncertainty with safe data-driven linear approximations of nonlinear dynamical systems.

The remainder of the paper is organized as follows. The proposed method is developed through Section \ref{sec3}. In Section \ref{sec5} we illustrate the proposed framework on a spring mass system and a power system application, an example of a multi-machine power system, and present a detailed analysis of the impact of synchrophasor measurement uncertainties on Koopman operator-based dynamic characterization. Section \ref{sec6} concludes the paper and outlines future work. 

\section{Method}\label{sec3}

\subsection{Notation}
Throughout the paper, we used italic bold letters to represent the random matrix and italic unbold letters to represent its random elements.
The vector $\mathbf{e}_{j}$ represents an $n$-dimensional vector with its $j$-element equal to one, and zero otherwise. $|\bm{A}|$ denotes the determinant of $\bm{A}$.

\subsection{Preliminaries}
{Numerous algorithms have been proposed to approximate the Koopman operator from data \cite{budivsic2012applied, tu2013dynamic,rowley2009spectral}, with DMD \cite{schmid2010dynamic} and extended DMD \cite{williams2015data} being prominent examples. This paper focuses on the algorithmic steps of DMD, a method that analyzes system dynamics by extracting dominant modes and their associated temporal dynamics from snapshot data.}

Suppose we have $m+1$ data snapshots $\bm{x}_{1},\hdots,\bm{x}_{m+1}$ sampled from a continuous-time system at instances $t_{1},\hdots,t_{m+1}$. Let us organize $\bm{X}=[\bm{x}_{1},\hdots,\bm{x}_{m}]$ and $\bm{Y}=[\bm{x}_{2},\hdots,\bm{x}_{m+1}]$. An estimate of the DMD operator, assuming a full-state observable, is given by
\begin{equation}
\bm{A}\approx \bm{X}^{\dag}\bm{Y}, \label{eq.5}
\end{equation} 

\noindent
where $\bm{X}^{\dag}$ is the Moore-Penrose pseudoinverse of $\bm{X}$. The elements of $\bm{X}$ and $\bm{Y}$ are assumed to follow a normal distribution $\mathcal{N}(\mu_{x},{\sigma_{x}^{2}})$ and $\mathcal{N}(\mu_{y},{\sigma_{y}^{2}})$, respectively.

{
\subsection{Sensitivity of DMD method to the Homoscedastic and Heteroscedastic Noise}
To estimate \(\bm{A}\), DMD methods often use a least squares approach \cite{Abolmasoumi2022} formulated as \(\bm{y}_{i} = \bm{A}\bm{x}_{i} + \bm{\varepsilon}_{i}\), where the errors \(\bm{\varepsilon}_{i}\) follow a normal distribution, \(\bm{\varepsilon}_{i} \sim \mathcal{N}(\mathbf{0},\bm{\Sigma}_{i})\). To account for the measurement uncertainty in the states, we characterize the covariance matrix of the errors as \(\bm{\Sigma}_{i} = \text{diag}(\sigma_1^{2}, \dots, \sigma_n^{2})\), capturing the variances.}
{
Incorporating this uncertainty into the least squares objective function gives:
\[
f(\bm{A}) = \sum_{i=1}^{m} \bm{r}_{i}^{\top} \bm{\Sigma}_{i}^{-1} \bm{r}_{i},
\]
where the residuals are \(\bm{r}_{i} = \bm{y}_{i} - \bm{A}\bm{x}_{i}\). To estimate \(\bm{A}\), we differentiate the objective function with respect to \(\bm{A}\) and set it to zero:
\[
\frac{\partial f(\bm{A})}{\partial \bm{A}} = 0 \quad \Rightarrow \quad \sum_{i=1}^{m} \bm{x}_{i}^{\top} \bm{\Sigma}_{i}^{-1} (\bm{y}_{i} - \bm{A}\bm{x}_{i}) = 0.
\]
In the case of homoscedastic noise, where the noise variance is the same for all instances \(\bm{\Sigma}_{1} = \bm{\Sigma}_{2} = \dots = \bm{\Sigma}_{m} = \bm{\Sigma}_{\bm{\varepsilon}}\), we can simplify the estimation as:
\[
\bm{\Sigma}_{\bm{\varepsilon}}^{-1} \sum_{i=1}^{m} \bm{x}_{i}^{\top}(\bm{y}_{i} - \bm{A} \bm{x}_{i})  = 0,
\]
leading to the standard DMD estimate in \eqref{eq.5}.}
{
However, in the case of heteroscedastic noise, where the noise variances differ across instances \(\bm{\Sigma}_{1} \neq \bm{\Sigma}_{2} \neq \dots \neq \bm{\Sigma}_{m}\), this variation cannot be handled directly by the DMD algorithm, as it attributes same confidence to each measurement. The presence of heteroscedastic noise significantly impacts DMD estimations, requiring advanced approaches to accurately account for the differing uncertainties.}

This paper devises a numerical method to quantify the impact of measurement uncertainty on the random variables $a_{ij}$, elements of $\bm{A}$. That is, it provides expectations and confidence bounds on $a_{ij}$. For that, we propagate the first and second moments of $\bm{X}^{\dag}$ and $\bm{Y}$. Although the moments of $\bm{Y}$ are straightforward, $\mathbb{E}[y_{kt}]=\mu_{y_{kt}}$ and $\mathbb{E}[y_{kt}^{2}]= \sigma^{2}_{y_{kt}}$, we need to derive the first and second moments of $x^{\downharpoonright}_{tk}$, elements of $\bm{X}^{\dag}$. This is done in the next subsection, followed by the moments of $a_{ij}$, the elements of $\bm{A}$.


\subsection{Element-wise moments of the DMD operator}
The Moore-Penrose pseudoinverse of $\bm{X}$ is given by
\begin{equation}\label{eq1}
\bm{X}^{\dag}=(\bm{X}^{\top}\bm{X})^{-1}\bm{X}^{\top},
\end{equation}

\noindent
and the $tk$ element of $\bm{X}^{\dag}$, $x_{tk}$, can be expressed as
\begin{equation}\label{7}
x_{tk}^{\downharpoonright}=\mathbf{e}_{k}^{\top}(\bm{X}\bm{X}^{\top})^{-1}\bm{x}_{t}.
\end{equation}

Now, let
\begin{equation}
 \bm{X}\bm{X}^{\top} =\bm{V}+\bm{x}_{t}\bm{x}_{t}^{\top},
\end{equation}
where $\bm{V}=\sum_{l=1,l\neq t}^{m}\bm{x}_{l}\bm{x}_{l}^{\top}$. Using the Sherman–Morrison formula \cite{sherman1950adjustment}, we get

\begin{equation}\label{9}
 (\bm{X}\bm{X}^{\top})^{-1}=\bm{V}^{-1}-\frac{\bm{V}^{-1}\bm{x}_{t}\bm{x}_{t}^{\top}\bm{V}^{-1}}{1+\bm{x}_{t}^{\top}\bm{V}^{-1}\bm{x}_{t}}.
\end{equation}

\noindent
where $\bm{V}^{-1}$ is symmetric. Substituting \eqref{9} into \eqref{7},

\begin{equation}
x_{tk}^{\downharpoonright}
=\mathbf{e}_{k}^{\top} \left( \bm{V}^{-1}-\frac{\bm{V}^{-1}\bm{x}_{t}\bm{x}_{t}^{\top}\bm{V}^{-1}}{1+\bm{x}_{t}^{\top}\bm{V}^{-1}\bm{x}_{t}} \right) \bm{x}_{t},
\end{equation}

\noindent
where $\kappa = (\bm{x}_{t}^{\top}\bm{V}^{-1}\bm{x}_{t})$ is a scalar, and

\begin{align}
x_{tk}^{\downharpoonright}
&=\mathbf{e}_{k}^{\top} \left( \frac{\bm{V}^{-1} +\bm{V}^{-1}\kappa -\bm{V}^{-1}\bm{x}_{t}\bm{x}_{t}^{\top}\bm{V}^{-1}}{1+\kappa} \right) \bm{x}_{t} \nonumber \\
&= \frac{\mathbf{e}_{k}^{\top}\bm{V}^{-1}\bm{x}_{t} +\mathbf{e}_{k}^{\top}\bm{V}^{-1}\kappa\bm{x}_{t} -\mathbf{e}_{k}^{\top}\bm{V}^{-1}\bm{x}_{t}\kappa}{1+\kappa} \nonumber \\
&=\frac{\mathbf{e}_{k}^{\top} \bm{V}^{-1}\bm{x}_{t}}{1+\bm{x}_{t}^{\top} \bm{V}^{-1}\bm{x}_{t}}. \label{11}
\end{align}

Now, for the sake of simplicity of notation, let $\bm{R}=\bm{V}^{-1}$. Let us also define $s_{1}=\mathbf{e}_{k}^{\top} \bm{R}\bm{x}_{t}$ and $s_{2}=1+\bm{x}_{t}^{\top} \bm{R}\bm{x}_{t}$.
In what follows, we derive the first and second moments of $x_{tk}^{\downharpoonright}$ using moment generating functions (MGFs). 
The MGF of a real-valued random variable is an alternative specification of its probability distribution; it encodes all the moments of a random variable into a single function from which they can be extracted again later. An MGF $h:\mathbb{R}\to [0,\infty)$ of a random variable $s$ and parameter $p$ is defined as
\begin{equation}
h(p) \coloneqq \mathbb{E}[\textnormal{exp}(p\cdot s)].
\end{equation}
The Taylor series expansion of the MGF around \( p = 0 \) is given by
\begin{equation*} 
h(p) = \mathbb{E}[\textnormal{exp}(p\cdot s)] = 1 + p \mathbb{E}[s] + \frac{p^2 \mathbb{E}[s^2]}{2!} + \frac{p^3 \mathbb{E}[s^3]}{3!} + \cdots
\end{equation*}
The \( \eta \)-th moment about 0 is the \( \eta \)-th derivative of the MGF evaluating at \( p = 0 \) is expressed as
\[
\mathbb{E}[s^\eta] = \frac{d^\eta}{dp^\eta} h(p) \bigg|_{p=0}.
\]

Note that $s_{1}$ and $s_{2}$ in \eqref{11} are both functions of random variables, thus requiring the joint MGF. Following Hoque \cite{hoque1985exact}, the joint MGF for a rational function with quadratic forms in the nominator and denominator is given by
\begin{align}\label{13}
h(p_{1},p_{2})=\mathbb{E}\left[\textnormal{exp}(p_{1}\cdot s_{1} +p_{2}\cdot s_{2})\right],
\end{align}

\noindent
where $p_{1}$ and $p_{2}$ are parameters. However, unlike \cite{hoque1985exact} where the matrix that multiplies the vector of random variables is deterministic, here $\bm{R}$ is stochastic. Therefore, we derive the joint MGF conditioned on $\bm{R}$ for the expression in equation $\eqref{11}$. Hence, we redefine the joint MGF in \eqref{13} is as follows:
\begin{align}\label{14}
h(p_{1},p_{2}|\bm{R})=\mathbb{E}\left[\textnormal{exp}(p_{1}\cdot s_{1}+p_{2}\cdot s_{2})|\bm{R}\right].
\end{align}
\begin{figure*}[b]
\noindent\makebox[\linewidth]{\rule{\textwidth}{0.4pt}}
\setcounter{equation}{12}
\begin{align}
 &h(p_1,-p_2|\bm{R})\notag
 =\int_{-\infty}^{+\infty} f(\bm{x})\;\textnormal{exp}(p_{1}\cdot s_{1}-p_{2}\cdot s_{2})d \bm{x}\notag\\
 &=\int_{-\infty}^{+\infty}\frac{1}{(2\pi)^{n/2}|\mathbf{\Sigma|}^{1/2}}\textnormal{exp}\left(-\frac{1}{2}(\bm{x}-{\bm{\mu}_{\bm{x}}})^{\top}\bm{\Sigma}^{-1}(\bm{x}-{\bm{\mu}_{\bm{x}}})\right) \textnormal{exp}\left(p_{1}\mathbf{e}^{\top}\bm{R}\bm{x}-p_{2}(1+\bm{x}^{\top}\bm{R}\bm{x})\right)d\bm{x}\notag\\
 &= \frac{1}{(2\pi)^{n/2}|\mathbf{\Sigma|}^{1/2}}\textnormal{exp}\left( -\frac{1}{2}{\bm{\mu}_{x}^{\top}\mathbf{\Sigma}^{-1}\bm{\mu}_{x}}\right) \int_{-\infty}^{+\infty} \textnormal{exp}\left( -\frac{1}{2} \bm{x}^{\top}\mathbf{\Sigma}^{-1}\bm{x} + \bm{x}^{\top}\mathbf{\Sigma}^{-1}{\bm{\mu}_{\bm{x}}}+p_{1}\mathbf{e}^{\top}\bm{R}\bm{x}-p_{2}-
 p_{2}\bm{x}^{\top}\bm{R}\bm{x}\right) d \bm{x} \notag\\
 &= c_{1} \int_{-\infty}^{+\infty} \textnormal{exp} \left( - \bm{x}^{\top}\Bigl(\frac{1}{2}\bm{\Sigma}^{-1}+p_{2}\bm{R}\Bigr)\bm{x} + (\bm{\Sigma}^{-1}{\bm{\mu}_{\bm{x}}}+
 p_{1}\bm{r})^{\top}\bm{x} \right) d \bm{x}
 =c_{1}\int_{-\infty}^{+\infty} \textnormal{exp}\left(-\bm{x}^{\top}\mathbf{S}\bm{x}+ (\mathbf{b}+p_{1}\bm{r})^{\top}\bm{x}\right) d \bm{x} \label{mgf}
\end{align}
\setcounter{equation}{13}
\end{figure*}
Now, drawing parallels to the approach in \cite{hoque1985exact}, we apply the integrals by Sawa \cite{sawa1978exact} to derive exact moments of the joint MGF. {Note that $h{(p_{1},-p_{2}|\bm{R})}$ incorporates the negative sign on $p_{2}$ to achieve the quadratic form of $s_{2}=1+\bm{x}_{t}^{\top}\bm{V}^{-1}\bm{x}_{t}$ in the denominator of the expression for the moments in \eqref{11}}.
Then, we have:
\setcounter{equation}{10}
\begin{align}\label{sawa}
&\mathbb{E}\left[({x_{tk}^{\downharpoonright}})^{\eta}\right] \nonumber \\
&=
\frac{1}{(\eta-1)!} \int_{0}^{\infty}p_{2}^{\eta-1} \left. \left(\frac{\partial^{\eta}}{\partial p_{1}^{\eta}} h(p_{1},-p_{2}|\bm{R}) \right)\right|_{p_{1}=0} dp_{2},
\end{align}
\setcounter{equation}{11}

to extract the $\eta^{\textrm{th}}$ moment of ${x_{tk}^{\downharpoonright}}$ from \eqref{14}, where $\eta$ is a positive integer.



Now we are in a position to state the main result of this paper.

\begin{theorem}
Let the $t$-column vector $\bm{x}_{t}$ of the data matrix $\bm{X}$ follow a multivariate normal distribution $\mathcal{N}(\bm{\mu_{x}},\bm{\Sigma})$. The exact first moment of $x^{\downharpoonright}_{tk}$, $k$-element of the $t$-column vector $\bm{x}^{\downharpoonright}_{t}$ of $\bm{X}^{\dag}$, the Moore-Penrose pseudoinverse of $\bm{X}$, is given by
\begin{align}\label{16}
 \mathbb{E}[{x_{tk}^{\downharpoonright}}]
 &=\mathbb{E}\left[\textnormal{exp}(p_{1} s_{1} -p_{2} s_{2})| \bm{R}\right]\notag\\
 &= c \int_{0}^{\infty} |\mathbf{S}|^{-1/2} \;\textnormal{exp}(-p_{2})\; \notag\\&\qquad\qquad\textnormal{exp}\left(\frac{\mathbf{b}^{\top}\mathbf{S}^{-1}\mathbf{b}}{4}\right) \cdot \left(\frac{\bm{r}^{\top}\mathbf{S}^{-1}\mathbf{b}}{2}\right) dp_{2},
\end{align}
where $\mathbf{S}=\frac{1}{2}\mathbf{\Sigma}^{-1}+p_{2}\bm{R}$, $\mathbf{b}=\mathbf{\Sigma}^{-1}\bm{\mu}_{\bm{x}}$,
$\bm{R}$ is symmetric with $k$-row ($k$-column) vector $\bm{r}^{\top}$ ($\bm{r}$), and
\begin{equation*}
c=\frac{1}{2^{n/2}|\mathbf{\Sigma|}^{1/2}} \textnormal{exp}\left(-\frac{1}{2}{\bm{\mu}_{\bm{x}}}^{\top}\bm{\Sigma}^{-1}{\bm{\mu}_{\bm{x}}}\right).
\end{equation*}
\end{theorem}

\begin{proof}
To simplify the notation, we omit the subscripts in $\bm{x}$ and $\mathbf{e}$. We develop MGF $h(p_{1},-p_{2}|\bm{R})$
in \eqref{mgf}, where $c_{1}=\frac{1}{(2\pi)^{n/2}|\mathbf{\Sigma|}^{1/2}} \textnormal{exp}(-\frac{1}{2}{\bm{\mu}_{\bm{x}}^{\top}}\bm{\Sigma}^{-1}{\bm{\mu}_{\bm{x}}}-p_{2})$.
\setcounter{equation}{13}
Using the Gaussian integral \cite{robert21} in the form
\begin{equation*}
 \int \textnormal{exp} \left(-\bm{u}^{\top}\bm{L}\bm{u}+\bm{v}^{\top}\bm{u} \right) d \bm{u} = \frac{\pi^{n/2}}{|\bm{L}|^{1/2}}\textnormal{exp}\left( \frac{\bm{v}^{\top}\bm{L}^{-1}\bm{v}}{4} \right),
\end{equation*}
we arrive at the solution of the integral in MGF \eqref{mgf}:
 \begin{equation}
 h(p_{1},-p_{2}|\bm{R})=c_{2}\;\textnormal{exp}\left(\frac{(\mathbf{b}+p_{1}\bm{r})^{\top}\mathbf{S}^{-1}(\mathbf{b}+p_{1}\bm{r})}{4}\right), \label{eq.18}
 \end{equation}
where $c_{2}=\frac{1}{2^{n/2}|\mathbf{S}|^{1/2}|\mathbf{\Sigma}|^{1/2}}\textnormal{exp}(-\frac{1}{2}{\bm{\mu}_{\bm{x}}^{\top}}\bm{\Sigma}^{-1}{\bm{\mu}_{\bm{x}}}-p_{2})$.

The first derivative of \eqref{eq.18} with respect to $p_{1}$, required to calculate the first moment of $x_{tk}^{\downharpoonright}$, is given by \eqref{17}. Evaluating \eqref{17} at $p_{1}=0$ yields
\setcounter{equation}{15}
\begin{small}
\begin{equation}\label{19}
\left. \frac{\partial}{\partial p_{1}} h(p_{1},-p_{2}|\bm{R}) \right|_{p_{1}=0} = c_{2}\; 
\textnormal{exp}\left(\frac{\mathbf{b}^{\top}\mathbf{S}^{-1}\mathbf{b}}{4}\right) \cdot\left(\frac{\bm{r}^{\top}\mathbf{S}^{-1}\mathbf{b}}{2}\right)
\end{equation}
\end{small}Substituting \eqref{19} into \eqref{sawa},
\begin{align*}
 \mathbb{E}[{x_{tk}^{\downharpoonright}}] &= c \int_{0}^{\infty} |\mathbf{S}|^{-1/2} \;\textnormal{exp}(-p_{2}) \notag\\
&\qquad \textnormal{exp}\left(\frac{\mathbf{b}^{\top}\mathbf{S}^{-1}\mathbf{b}}{4}\right) \cdot \left(\frac{\bm{r}^{\top}\mathbf{S}^{-1}\mathbf{b}}{2}\right) dp_{2}.
\end{align*}
\end{proof}
\setcounter{equation}{14}
\begin{figure*}[t]
\begin{align}\label{17}
&\frac{\partial h(p_{1},-p_{2}|\bm{R})}{\partial p_{1}} ={C}_{2}\; \textnormal{exp}\left(\frac{\mathbf{b}^{\top}\mathbf{S}^{-1}\mathbf{b}}{4}\right)\textnormal{exp}\left(\frac{p_{1}\bm{r}^{\top}\mathbf{S}^{-1}\mathbf{b}}{2}+\frac{p_{1}^{2}\bm{r}^{\top}\mathbf{S}^{-1}\bm{r}}{4}\right)\cdot\left(\frac{\bm{r}^{\top}\mathbf{S}^{-1}\mathbf{b}}{2}+\frac{p_{1}\bm{r}^{\top}\mathbf{S}^{-1}\bm{r}}{2} \right)
\end{align}
\end{figure*}
\setcounter{equation}{17}
\begin{figure*}[t]
\vspace{-.5cm}
\begin{small}
\begin{align}
 &\frac{\partial^{2}h(p_{1},-p_{2}|\bm{R})}{\partial p_{1}^{2}}=c_{3} \;\textnormal{exp}\left( \frac{p_{1}\bm{r}^{\top}\mathbf{S}^{-1}\mathbf{b}}{2} +\frac {p_{1}^{2}\bm{r}^{\top}\mathbf{S}^{-1}\bm{r}}{4}\right)
 \left[\frac{\left(\bm{r}^{\top}\bm{S}^{-1}\bm{b}\right)}{4} \left(\bm{r}^{\top}\bm{S}^{-1}\bm{b} + 2p_{1}\bm{r}^{\top}\bm{S}^{-1}\bm{r}\right) + \frac{\left(\bm{r}^{\top}\bm{S}^{-1}\bm{r}\right)}{4} \left(p_{1}^{2}\bm{r}^{\top}\bm{S}^{-1}\bm{r} + 2\right) \right] \label{eq.22}
\end{align}
\end{small}
\noindent\makebox[\linewidth]{\rule{\textwidth}{0.4pt}}
\end{figure*}
\setcounter{equation}{18}

\begin{theorem}
Let the $t$-column vector $\bm{x}_{t}$ of the data matrix $\bm{X}$ follow a multivariate normal distribution $\mathcal{N}(\bm{\mu_{x}},\bm{\Sigma})$. The exact second moment of $x^{\downharpoonright}_{tk}$, $k$-element of the $t$-column vector $\bm{x}^{\downharpoonright}_{t}$ of $\bm{X}^{\dag}$, the Moore-Penrose pseudoinverse of $\bm{X}$, is given by
\setcounter{equation}{16}
\begin{align}\label{22}
&\mathbb{E}\left[({x_{tk}^{\downharpoonright}})^{2}\right] = c\; \int_{0}^{\infty} p_{2}|\mathbf{S}|^{-1/2} \;\textnormal{exp}(-p_{2}) \notag \\
 &\qquad \textnormal{exp}\left( \frac{\mathbf{b}^{\top}\mathbf{S}^{-1}\mathbf{b}}{4}\right) \cdot \left(\frac{\bm{r}^{\top}\mathbf{S}^{-1}\bm{r}}{2} + \frac{(\bm{r}^{\top}\mathbf{S}^{-1}\mathbf{b})^{2}}{4} \right) dp_{2}.
\end{align}

\noindent
\end{theorem}
\begin{proof}Starting from \eqref{17}, we obtain \eqref{eq.22}, where $c_{3}=c_{2}\cdot \textnormal{exp}\left(\frac{\mathbf{b}^{\top}\mathbf{S}^{-1}\mathbf{b}}{4}\right)$. Evaluating \eqref{eq.22} at $p_{1}=0$,

\setcounter{equation}{18}
\begin{equation}\label{20}
\left.\frac{\partial^{2}}{\partial p_{1}^{2}} h(p_{1},-p_{2}|\bm{R}) \right|_{p
_{1}=0}= c_{3} \left( \frac{(\bm{r}^{\top}\mathbf{S}^{-1}\mathbf{b})^{2}}{4} + \frac{\bm{r}^{\top}\mathbf{S}^{-1}\bm{r}}{2}\right)
\end{equation}
and substituting \eqref{20} into \eqref{sawa} yields \eqref{22}.
\end{proof}

{Although we assumed independence among the variables in \eqref{mgf} by using a diagonal covariance matrix \(\bm{\Sigma}\), the correlations between the variables in \(\bm{x}_{t}\) can be introduced by including off-diagonal elements in the covariance matrix.}

We proceed with an element-wise derivation of the first and second moments of $\bm{A}$ in \eqref{eq.5}. Recall that our objective is to quantify the impact of measurement uncertainty on the random variables $a_{ij}$, elements of $\bm{A}$. Let
\begin{equation}\label{1}
 {a}_{ij}= {x}_{i1}^{\downharpoonright}{y}_{1j}+\hdots+{x}_{in}^{\downharpoonright}{y}_{nj} = \sum_{k=1}^{n} {x}_{ik}^{\downharpoonright}{y}_{kj},
\end{equation}
$i=1,\hdots,m$ and $j=1,\hdots,m$. The first moment of $a_{ij}$ is
\begin{align}
\mathbb{E}\left[{a}_{ij}\right] &=
\sum_{k=1}^{n} \mathbb{E}\left[{x}_{ik}^{\downharpoonright}{y}_{kj}\right] = \sum_{k=1}^{n} \mathbb{E}\left[{x}_{ik}^{\downharpoonright}\right]\mathbb{E}\left[{y}_{kj}\right]\notag\\
&= \sum_{k=1}^{n} \mathbb{E}\left[{x}_{ik}^{\downharpoonright}\right] \mu_{{y}_{kj}},
\end{align}

\noindent
where $\mu_{{y}_{kj}}$ is straightforward to estimate using the sample mean. On the other hand, we use the result from Theorem 1 to estimate $\mathbb{E}\left[{x}_{ik}^{\downharpoonright}\right]$. 
Let us gather the estimates of $\mathbb{E}\left[{x}_{ik}^{\downharpoonright}\right]$ in a matrix $\bm{M}^{(1)}_{\bm{X}^{\dag}}$, and denote its $i$-row by $\bm{\mu}^{\top}_{\bm{x}_{i}^{\downharpoonright}}$. Similarly, we gather the estimated values of $\mu_{{y}_{kj}}$ in a matrix $\bm{M}^{(1)}_{\bm{Y}}$ and denote its $j$-column by $\bm{\mu}_{\bm{y}_{j}}$. Then, an estimate of the first moment of $a_{ij}$ is given by:
\begin{equation}\label{m1}
\widehat{\mu}_{a_{ij}}=\bm{\mu}^{\top}_{\bm{x}_{i}^{\downharpoonright}}\bm{\mu}_{\bm{y}_{j}}.
\end{equation}

We now focus on obtaining an expression for the element-wise second moments of $\bm{A}$ in \eqref{eq.5}. To this aim, starting from \eqref{1},
 \begin{align}\label{29}
\mathbb{E}[{a}_{ij}^{2}] &= \mathbb{E}\left[(x_{i1}^{\downharpoonright} y_{1j} + \hdots+ x_{in}^{\downharpoonright} y_{nj})^{2}\right]\notag\\ 
&= \mathbb{E}\left[( x_{i1}^{\downharpoonright} y_{1j})^{2}\right] - \mathbb{E}\left[x_{i1}^{\downharpoonright}\right]^{2} \mathbb{E}\left[y_{1j}\right]^{2}+\hdots \notag\\
&\qquad+ \mathbb{E}\left[(x_{in}^{\downharpoonright} y_{nj})^{2}\right] -\mathbb{E}\left[x_{in}^{\downharpoonright}\right]^{2} \mathbb{E}\left[y_{nj}\right]^{2}\notag\\
&= \mathbb{E}\left[(x^{\downharpoonright}_{i1})^{2}\right] \mathbb{E}\left[y_{1j}^2\right] - \mathbb{E}\left[x_{i1}^{\downharpoonright}\right]^{2} \mathbb{E}\left[y_{1j}\right]^{2} + \hdots \notag \\ 
&\qquad+\mathbb{E}\left[(x^{\downharpoonright}_{in})^{2}\right] \mathbb{E}\left[y_{nj}^2\right] -\mathbb{E}\left[x_{in}^{\downharpoonright}\right]^{2} \mathbb{E}\left[y_{nj}\right]^{2}\notag\\
&=\sum_{k=1}^{n}\left(\mathbb{E}\left[(x^{\downharpoonright}_{ik})^{2}\right]\mathbb{E}\left[y_{kj}^2\right] -\mathbb{E}\left[x_{ik}^{\downharpoonright}\right]^{2}\mathbb{E}\left[y_{kj}\right]^{2}\right).
\end{align} 
\begin{algorithm}[t!]
\caption{Estimate the first and second moments of $\bm{A}$}\label{Al1}
\begin{algorithmic}
\State \textbullet\ Obtain the data $\mathcal{D}=\{\mathbf{X},\mathbf{Y}\}$

\State \textbullet\ Obtain or calculate $\bm{\sigma}^{2}_{\bm{y}}=[\sigma_{1}^{2},\hdots,\sigma_{n}^{2}]^{\top}$

\State \textbullet\ Apply \eqref{16} to estimate the first moments $\bm{M}^{(1)}_{\bm{X}^{\dag}}$

\State \textbullet\ Apply \eqref{22} to estimate the second moments $\bm{M}^{(2)}_{\bm{X}^{\dag}}$

\State \textbullet\ Initialize $i=1$ and $j=1$
\For {$i:1:m$} 
 \For {$j:1:m$} 
\State {$\boldsymbol{-}$} $m_{1}(i,j)=\bm{\mu}^{\top}_{\bm{x}_{i}^{\downharpoonright}}{\bm{\mu}_{\bm{y}_{j}}}$
\vspace{0.4em}
\State {$\boldsymbol{-}$} $m_{2}(i,j)=({\bm{\sigma}^{2}_{\bm{x}_{i}^{\downharpoonright}}})^{\top} 
(\bm{\sigma}_{\bm{y}}^{2})- (\bm{\mu}^{2}_{\bm{x}_{i}^{\downharpoonright}})^{\top}{(\bm{\mu}^{2}_{\bm{y}_{j}})}.$
\EndFor
\State \textbullet\ $\widehat{\mu}_{a_{ij}}=m_{1}(i,j)$
\State \textbullet\ $\widehat{\sigma}^{2}_{a_{ij}}=m_{2}(i,j)$
\EndFor
\end{algorithmic}
\end{algorithm}
Note that the uncertainty around the $k$-state $x_{k}$ is given by its associated variance, $\sigma^{2}_{k}$. In what follows, we assume the random variables to be \emph{homoscedastic}, that is, all random variables have the same finite variance. This is also known as homogeneity of variance. Thus, the random element $y_{kj}$ follows a normal distribution $\mathcal{N}(\mu_{{y}_{kj}}, \sigma^{2}_{k})$. 

Substituting $\mu_{{y}_{kj}}$ for the mean and $\sigma^{2}_{k}$ for the variance of $y_{kj}$ in \eqref{29}, the second moment of $a_{ij}$ is given by:
\begin{equation}
 \mathbb{E}\left[({a}_{ij})^{2}\right] = \sum_{k=1}^{n}\mathbb{E}\left[(x^{\downharpoonright}_{ik})^{2}\right]\sigma_{k}^2 -\mathbb{E}\left[x_{ik}^{\downharpoonright}\right]^{2}\mu_{{y}_{kj}}^{2}.
\end{equation} 

We now make use of the result in Theorem 2. Let us gather the estimates of $\mathbb{E}\left[(x^{\downharpoonright}_{ik})^{2}\right]$ in $\bm{M}^{(2)}_{\bm{X}^{\dag}}$, and denote its $i$-row by 
$({\bm{\sigma}^{2}_{\bm{x}_{i}^{\downharpoonright}}})^{\top}$. The confidence levels of the random elements $a_{ij}$, which represent the measurement uncertainties propagated from system states, can finally be estimated as 
\begin{equation}\label{m2}
\widehat{\sigma}^{2}_{a_{ij}}=({\bm{\sigma}^{2}_{\bm{x}_{i}^{\downharpoonright}}})^{\top} 
(\bm{\sigma}_{\bm{y}}^{2})- (\bm{\mu}^{2}_{\bm{x}_{i}^{\downharpoonright}})^{\top}{(\bm{\mu}^{2}_{\bm{y}_{j}})},
\end{equation}
where $\bm{\sigma}_{\bm{y}}^{2}=[\sigma_{1}^{2},\hdots,\sigma_{n}^{2}]^{\top}$.
Algorithm \ref{Al1} summarizes the process for estimating the confidence levels in terms of the first and second moments of the random elements of $\bm{A}$.

\begin{remark}
Note that our proposal to quantify the impact of measurement uncertainty on the elements of $\bm{A}$ is agnostic of specific dynamic mode decomposition methods. This feature is desirable, given the number of variants (see, e.g., \cite{Ichinaga2024}) of the original dynamic mode decomposition method \cite{Schmid2010}.
\end{remark}

\section{Numerical Simulations}\label{sec5}
\begin{figure}[!ht]
 \centering
 \includegraphics[width=8.9cm]{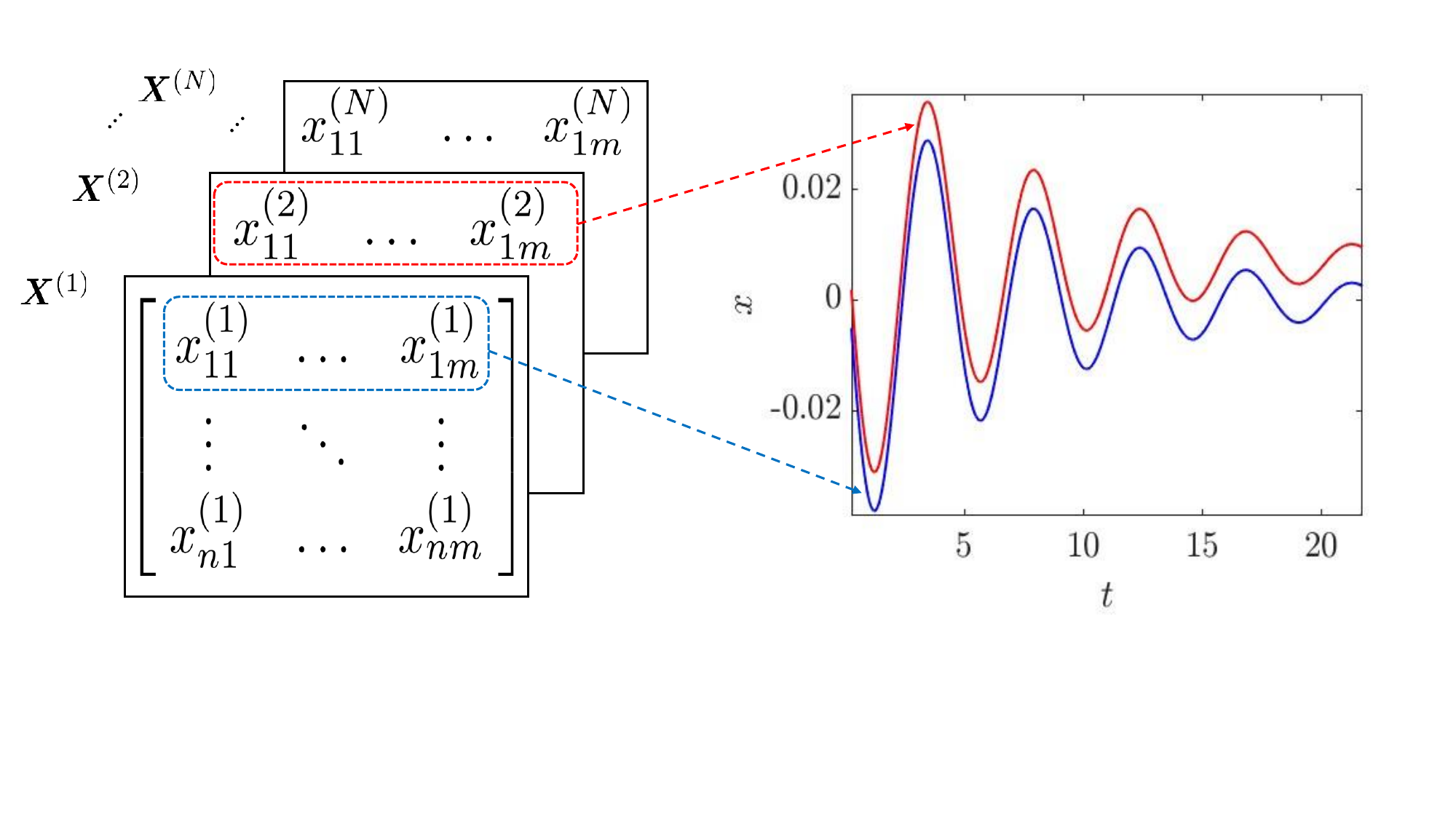}
 \vspace{-1.7cm}
 \caption{Obtaining instances of the random matrix $\bm{X}$ from the normal distribution of elements $x\sim\mathcal{N}(\mu_{x},\sigma^{2}_{x}).$ }
 \label{demo}
\end{figure}

We perform simulations on a spring-mass system and a multi-machine power system. The covariance matrix $\bm{\Sigma}$ captures the measurement uncertainty of the recorded states; the two ways to accomplish this are (i) to use the standard deviations of the measurement devices provided by the manufacturers and (ii) to calculate the variance of the ambient measurements using sample variance. We make the blanket assumption of a Gaussian distribution around the recorded data in each instance: $x_{tk}\sim \mathcal{N}(\textrm{x}_{tk},\sigma^{2}_{k})$ and $y_{tk}\sim \mathcal{N}(\textrm{y}_{tk},\sigma^{2}_{k})$. We employ \eqref{16} and \eqref{22} to estimate the element-wise mean and variance of $\bm{X}^{\dag}$, necessary for the DMD method. Following Algorithm \ref{Al1}, we further estimate the first and second moments of the elements of $\bm{A}$.
The estimated moments are compared with those obtained from Monte Carlo simulations, which are the true values. To perform Monte Carlo simulations, $N$ random trajectories of each state are drawn as illustrated in Figure \ref{demo}. Then, $N$ samples of $\bm{A}$ are obtained using DMD in \eqref{eq.5} as
\begin{equation}
    \bm{A}^{(l)}={\bm{X}^{\dag}}^{(l)}\bm{Y}^{(l)},
\end{equation}
$l=1,\hdots,N.$
Finally, the first and second moments of $a_{ij}$ are estimated using sample mean ${\mu}_{a_{ij}}=\frac{1}{N}\sum_{l=1}^{N}{a}_{ij}^{(l)}$ and sample variance ${\sigma}^{2}_{a_{ij}}=\frac{1}{N-1}\sum_{l=1}^{N}{({a}_{ij}^{{(l)}}-\textrm{a}_{{ij}})^{2}}$, respectively.

\subsection{Spring-Mass system}
\begin{figure}[!ht]%
\centering
\includegraphics[width=\linewidth]{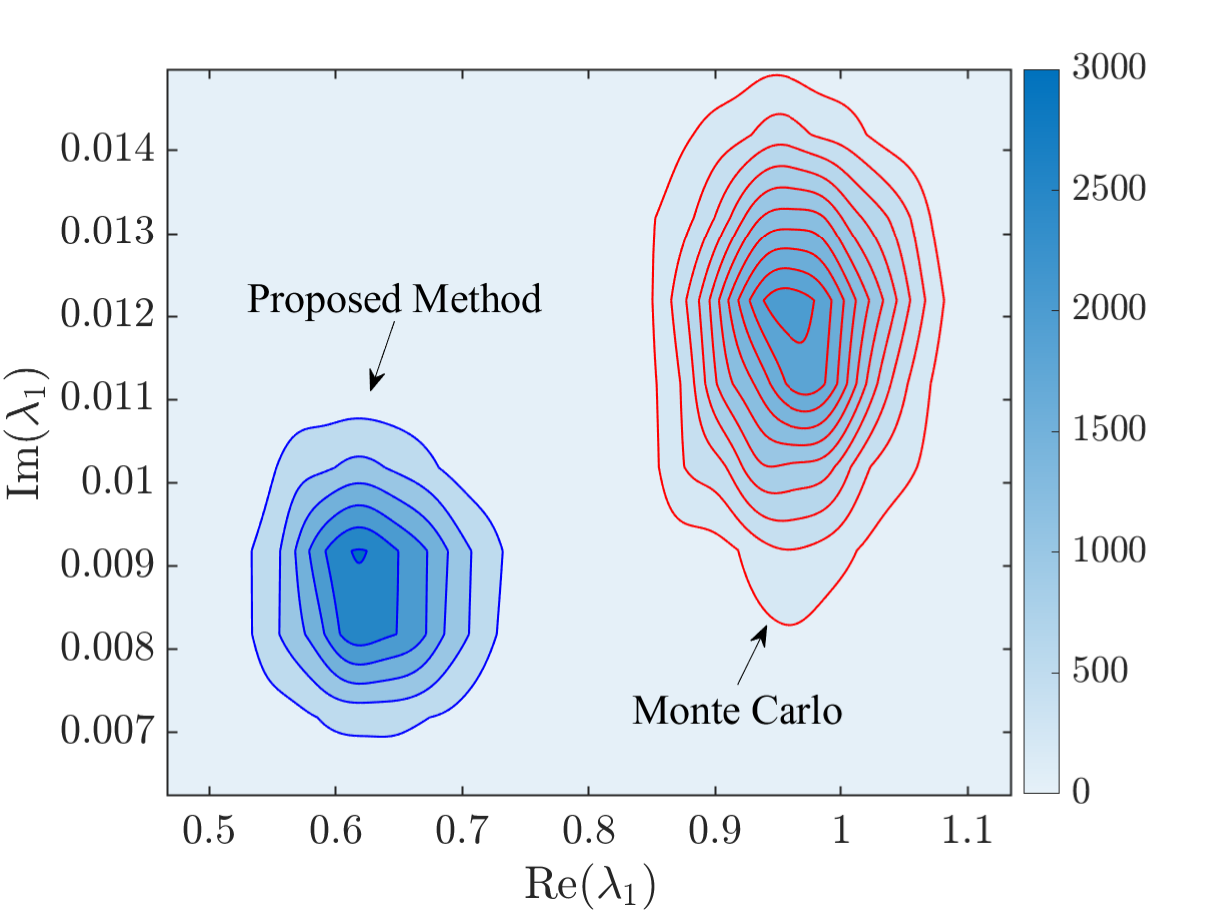}
\vspace{-0.6cm}
\caption{{Comparison of kernel densities for the largest eigenvalue $\lambda_1$ of $\bm{A}$: Density distributions estimated from samples obtained through our proposed method and Monte Carlo simulation. The colorbar indicates density values.}}%
\label{prob_2}%
\end{figure} 

Consider a spring-mass system with $n=2$ states: displacement $x_{1}$ and velocity $x_{2}$, as follows:
\begin{align*}
    \dot{x}_{1} &= x_{2}, \\
    \dot{x}_{2} &= \frac{-k}{m}x_{1}-g, 
\end{align*}
where $m$ is the mass of the object, $k$ is the spring constant, $g$ is the acceleration due to gravity. {We choose $m=5$ kg and $k=20$ N/m.}
The simulation with initial conditions set as $\bm{x}_{0}=[0.03;\,0.01]$ recorded for $40$ seconds constitutes our observed data $\mathcal{D}_{obs}=\{\mathbf{X},\mathbf{Y}\}$. We estimate $\sigma_{1}^{2}$ and $\sigma_{2}^{2}$ using the sample variance of the measurements taken between $30$--$40$ seconds, where they remain steady. The diagonal matrix $\bm{\Sigma}=\textrm{diag}(\sigma_{1}^{2},\sigma_{2}^{2})$ now constitutes the uncertainty in the states.

The estimated first (second) moments of $a_{ij}$ employing Algorithm \ref{Al1} are gathered in $\bm{M}^{(1)}_{\bm{A}}$ ($\bm{M}^{(2)}_{\bm{A}}$). The matrices $\bm{M}^{(1)}_{\bm{A}}$ and $\bm{M}^{(2)}_{\bm{A}}$ are compared with those obtained from Monte Carlo simulations.    
A similar comparison is made for the matrices $\bm{M}^{(1)}_{\bm{X}^{\dag}}$ and $\bm{M}^{(2)}_{\bm{X}^{\dag}}$ that constitute the first and the second moments of $x_{tk}^{\downharpoonright}$, respectively.
Table \ref{tab1} lists the root mean square error (RMSE), mean absolute error (MAE), Frobenius norm (Fr-norm), and cosine similarity (COS) of the performed comparisons. 

Let us obtain $N=1000$ instances $\bm{A}^{(l)};\;l=1,\hdots, N$ of $\bm{A}$ assuming a normal distribution on its elements $a_{ij}$ with means as their first moments and variances as their second moments estimated using the proposed method as
$a_{ij}^{(l)}\sim\mathcal{N}(\widehat{\mu}_{a_{ij}},\widehat{\sigma}^{2}_{a_{ij}})$.
For each instance $\bm{A}^{(l)}$, we calculate the eigenvalues $\{\lambda_{1}^{(l)}\geq\lambda_{2}^{(l)}\geq\hdots\lambda_{m}^{(l)} \}$ to obtain their $N$ samples. Similarly, let us obtain $N$ instances of eigenvalues of $\bm{A}^{(l)}$ generated using Monte Carlo simulation. Since the two largest eigenvalues $\{\lambda_{1},\lambda_{2}\}$ form a conjugate pair, we focus on examining the densities of the real and imaginary parts of the largest eigenvalue $\lambda_{1}$.
Figure \ref{prob_2} compares the kernel density of the largest eigenvalue $\lambda_{1}$ estimated from these samples obtained by both the proposed and Monte Carlo methods, with a squared exponential kernel fit. 

\subsection{Multi-machine Power System}\label{section_realdata}
\begin{figure}%
\centering
\subfloat[\centering ]{{\includegraphics[width=4.4cm]{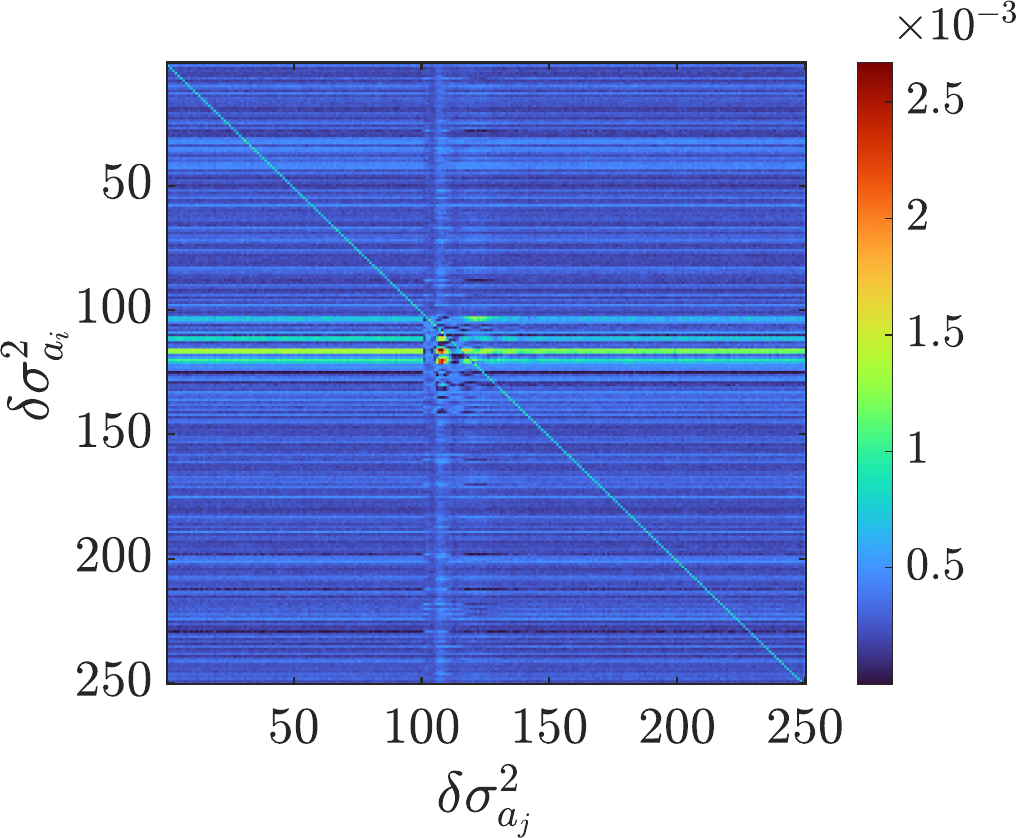}}}%
\subfloat[\centering ]{{\includegraphics[width=4.4cm]{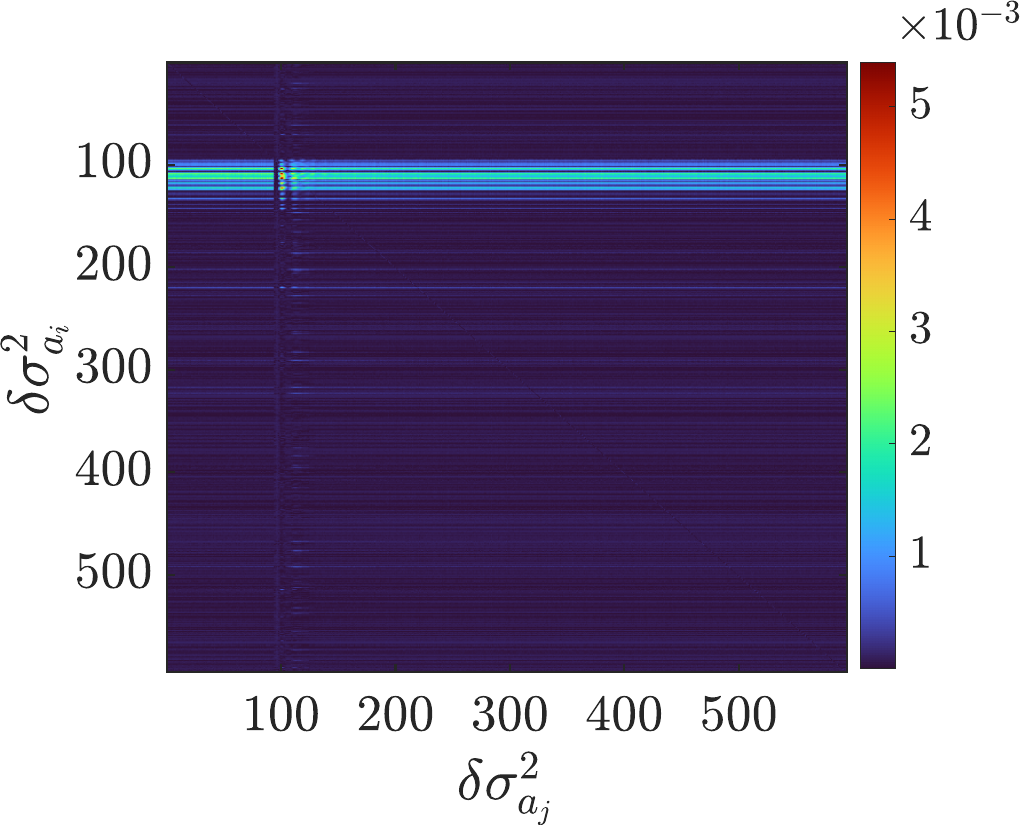}}}%
 \vspace{-0.2cm}
\caption{{Absolute error differences between the estimated $\widehat{\sigma}^{2}_{a_{ij}}$ and true second moments $\sigma^{2}_{a_{ij}}$ of DMD operator for
 (a) event A and (b) event B for the multi-machine power system, where ${\delta \sigma^{2}}_{a_{ij}}=|\sigma^{2}_{a_{ij}}-\widehat{\sigma}^{2}_{a_{ij}}|$.}}%
 \label{abs}
\end{figure}%
\begin{figure}
 \centering
\subfloat[\centering ]{{\includegraphics[width=8.8cm]{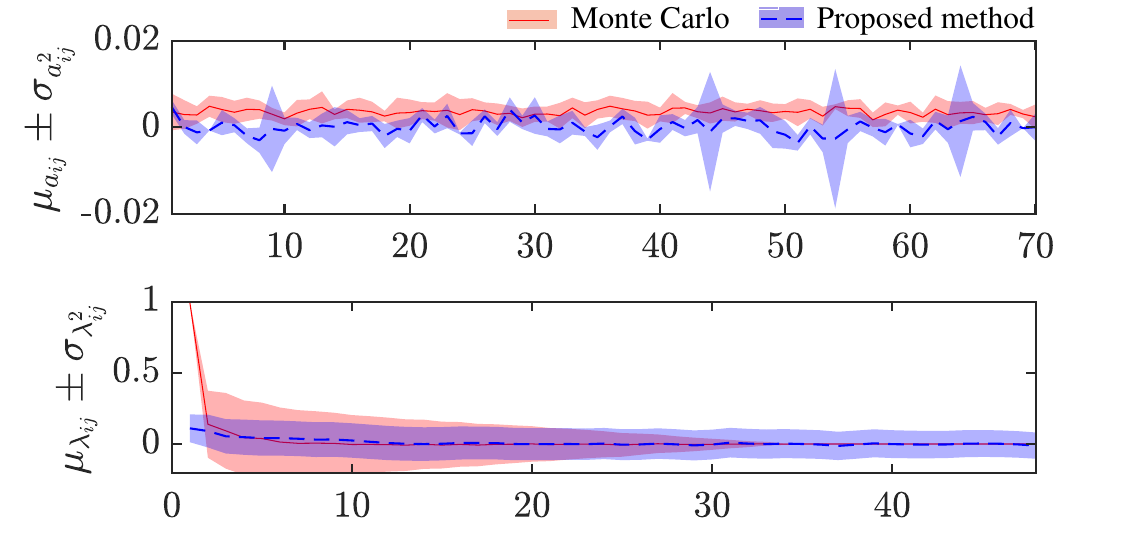}}}%
 \vspace{-0.4cm}
\quad
\subfloat[\centering ]{{\includegraphics[width=8.8cm]{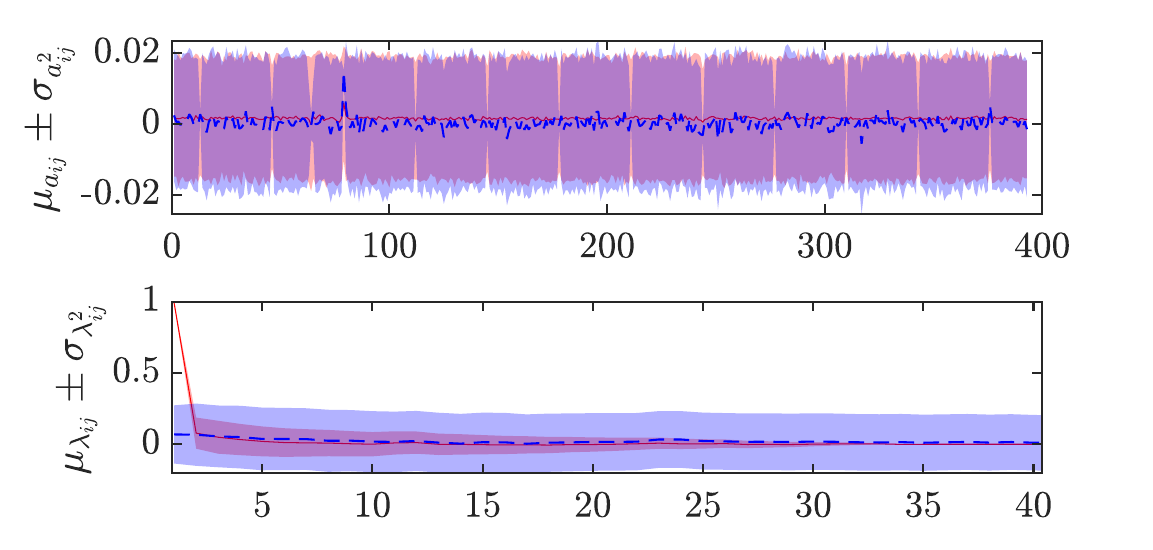}}}%
\vspace{-0.2cm}
 \caption{Comparison of kernel densities for DMD operator $a_{ij}$ and its eigenvalues $\lambda_{i}$ in the case of (a) even A and (b) event B estimated using samples obtained from the proposed method and Monte Carlo simulations for the multi-machine system.}
 \label{4mc}
\end{figure}
\begin{table*}[!ht]
    \centering
    \caption{The RMSE, MAE, Fr-norm, and COS values for the spring-mass system and multi-machine power system}
    \begin{tabular}{c|cccc|cccc|cccc}\hline
{} & \multicolumn{4}{c|}{Spring Mass System} & \multicolumn{8}{c}{Multi-machine Power System}\\
\cline{6-13}
{} & \multicolumn{4}{c|}{} & \multicolumn{4}{c|}{Event A} & \multicolumn{4}{c}{Event B}\\
\cline{2-13}
Measure & RMSE & MAE & Fr-norm & COS & RMSE & MAE & Fr-norm & COS & RMSE & MAE & Fr-norm & COS\\
\hline
$\bm{M}^{(1)}_{\bm{X}^{\dag}}$ & 3.79e-2&3.32e-2&1.19&1.24 & 4.94e{-2}&3.14e{-2}&4.55&1.86& 1.97e-2&1.19e-2&2.80&1.57\\
$\bm{M}^{(2)}_{\bm{X}^{\dag}}$ &2.45e-3&2.36e-3&7.75e-2&9.75e-1 & 2.82e^{-4}&9.15e-5&2.60e-2&5.22e-1& 3.39e-5&1.25e-5&4.83e-3&7.10e-1\\
$\bm{M}^{(1)}_{\bm{A}}$ & 1.04e-3&9.15e-4&5.23e-1&1.41 & 2.14e-2&1.58e-2&5.35&4.66 &8.14e-3&6.16e-3&4.83&4.88\\
$\bm{M}^{(2)}_{\bm{A}}$ &4.84e-6&4.48e-6&2.41e-3&9.17e-1 & 3.38e-4&3.04e-4&8.47e-2&7.28e-1 & 2.78e-4&1.11e-4&1.6e-1&5.92e-1\\ \hline
\end{tabular}
\label{tab1}
\end{table*}

The data obtained from {time domain simulation} of the multi-machine power system \cite{kundur2007power} 
is gathered in $\mathcal{D}_{obs}$. For this study, we considered detailed dynamical models associated with synchronous generators that led us to $n=34$ states in total simulated for $m= 120$ s. 
The dynamical events are further divided into two sub-events A and B,  with measurement uncertainty characterized by the steady-state period from 0 s to 53.99 s and from 63.18 s to 87.20 s, respectively. 

Table \ref{tab1} compares $\bm{M}^{(1)}_{\bm{X}^{\dag}}$,\;$\bm{M}^{(2)}_{\bm{X}^{\dag}}$, first and second moments of $\bm{X}^{\dag}$, and $\bm{M}^{(1)}_{\bm{A}}$,\;$\bm{M}^{(2)}_{\bm{A}}$, the first and second moments of $\bm{A}$ estimated using the proposed method with their true values obtained from Monte Carlo simulations. The comparison is conducted using RMSE, MAE, Fr-norm, and COS. Figure \ref{abs} illustrates the absolute error differences ${\delta \sigma^{2}}_{a_{ij}}=|\sigma^{2}_{a_{ij}}-\widehat{\sigma}^{2}_{a_{ij}}|$ between the true second moment ${{\sigma}}^{2}_{{a_{ij}}}$ obtained from Monte Carlo simulations and its estimated counterpart ${\widehat{\sigma}}^{2}_{{a_{ij}}}$ using the proposed method. 

In Figure \ref{4mc}, a detailed comparison of $\mu_{a_{ij}}$ and $\sigma^{2}_{a_{ij}}$, the first 
and the second moments of $\bm{A}$, estimated using the proposed method, are made with those obtained from Monte Carlo simulations.
Given the considerable number of elements of $\bm{A}$ (specifically $m \times m = 6.25e{4}\; (3.52e{5})$ in the case of event A (event B)), visualizing all of them does not help us to compare their estimated moments with the true ones. To address this, we plot the moments of $a_{ij}$ in the intervals of $900$ data points for better visualization. Additionally, to ensure consistency, the second moments are normalized using min-max scaling.
Figure \ref{4mc} also compares ${\mu}_{\lambda_{i}},{\sigma}^{2}_{\lambda_{i}}$, the first and second moments of the eigenvalues, estimated by employing sample mean and sample variance on $N$ samples obtained using the proposed method and Monte Carlo simulations as explained above, $i,j=1,\hdots,m$. 
The shaded region represents the range within $\pm 2$ second moments from the first moments. 

Remarkably, the absolute errors associated with the second moments of most of the DMD operator elements in Figure \ref{abs} are on the order of $10^{-3}$. The estimated first and second moments of $\bm{A}$ and its eigenvalues in Figure \ref{4mc} exhibit a high degree of comparability with the values obtained from Monte Carlo simulations. Similar is the case for the kernel densities of the eigenvalues compared in Figure \ref{prob_2}. Further strengthening the reliability of our approach, Table \ref{tab1} illustrates the accurate estimation of moments for $\bm{X}^{\dag}$. 
Accurate estimations of the first and second moments for $\bm{X}^{\dag}$, $\bm{A}$, and $\bm{\lambda}$ in our proposed measurement uncertainty analysis instill trustworthiness in the DMD method. 

\section{CONCLUSIONS}\label{sec6}
This study addresses the critical challenge of assessing uncertainties propagated from system measurements in the context of Koopman-theoretic data-driven characterization of dynamical systems, particularly through the lens of dynamic mode decomposition (DMD). Recognizing the substantial impact of data quality on the efficacy of DMD, we introduced a novel analytical approach for quantifying uncertainties in each constituent element of the approximated Koopman operator. The proposed method focuses on numerically estimating the confidence levels for the elements in terms of the first and second moments of the pseudo-inverse of the data matrix and the matrix of time-shifted snapshots of the data. Through detailed numerical analyses, we demonstrated the effectiveness of our approach in characterizing the dynamics of a spring-mass system and a multi-machine power system. This work contributes valuable insights into quantifying the local uncertainty of the data-driven characterization of nonlinear dynamical systems.

While we focus on the first and second moments in this paper, the formulation we present is general and one can follow the same ideas to derive higher-order moments. For example, the third moment, \emph{skewness}, indicates any \emph{asymmetric leaning} to either left or right and might be of interest within the context of data-driven operator learning. We leave that exploration for future work along with the derivation of the moments for non-normal cases.

\section*{ACKNOWLEDGEMENT} \label{sec7}
The authors would like to thank the U.S. Department of
Energy (DOE) for funding this research and Dr. Rui Yang for her support and guidance on this project. 

\end{document}